\documentclass[12pt,a4paper,notitlepage,oneside]{article}

\usepackage{alltt}
\usepackage[T1]{fontenc}
\usepackage[utf8]{inputenc}
\usepackage[english]{babel}

\usepackage{amsmath, amsthm, amssymb, mathtools}
\usepackage{dsfont}
\usepackage[cal=cm]{mathalfa}
\numberwithin{equation}{section}
\usepackage{chngcntr}
\counterwithout{equation}{section}

\usepackage{setspace} 
\usepackage{graphicx}
\usepackage{adjustbox}
\usepackage[usenames,dvipsnames]{xcolor}

\usepackage{float} 
\usepackage{wrapfig}
\usepackage{subcaption}
\usepackage[format=plain,labelfont=it,textfont=normalfont]{caption}

\usepackage{booktabs}

\usepackage{todonotes} 
\usepackage{url}
\usepackage{pdflscape} 
\usepackage{framed} 
\usepackage{enumitem} 

\usepackage[iso,swedish]{isodate}

\usepackage[round,colon,authoryear]{natbib}

\renewcommand{\P}{\mathbb{P}} 
\newcommand{\E}{\mathbb{E}} 
\newcommand{\V}{\mathbb{V}}
\newcommand{\bs}{\boldsymbol}
\newcommand{\IX}{\bs{1}}

\DeclareMathOperator*{\argmax}{arg\,max}

\usepackage{abstract_def}

\usepackage{titling}
\usepackage{authblk}
\title{An expectation-based space-time scan statistic for ZIP-distributed data}
\author[1]{Benjamin Allévius}
\author[1]{Michael Höhle}
\affil[1]{Department of Mathematics, Stockholm University, Sweden}
\date{\vspace{-5ex}}

\begin{document}
\maketitle

\begin{chapabstract} 
  \small{
\noindent 
An expectation-based scan statistic is proposed for the prospective 
monitoring of spatio-temporal count data with an excess of zeros. The method, 
which is based on an outbreak model for the
zero-inflated Poisson distribution, is shown to be superior to traditional scan
statistics based on the Poisson distribution in the presence of structural 
zeros. The spatial accuracy and detection timeliness of the proposed scan 
statistic is investigated by means of simulation, and an application on weekly 
cases of Campylobacteriosis in Germany illustrates how the scan statistic could 
be used to detect emerging disease outbreaks. An implementation of the method is 
provided in the open source \textsc{R} package \verb|scanstatistics| available
on CRAN. \\

\noindent
\textbf{Keywords}: EM algorithm, disease surveillance, scan statistic, 
                    spatio-temporal, zero-inflated Poisson.
}
  
\end{chapabstract}
\thispagestyle{empty}
\thispagestyle{empty}
\clearpage
\setcounter{page}{1}
\section{Introduction} \label{sec:intro}
The resurgence of infectious diseases such as Ebola and Zika in recent years has 
put the efforts of public health agencies in the spotlight. These agencies often 
monitor several hundred different pathogens as they are reported by local 
clinics and hospitals across a country. The movement of both 
people and foodstuff within and between countries enables diseases to spread 
quickly and can cause outbreaks that affect many locations simultaneously. By 
the joint surveillance of data from multiple geographical areas, be they 
different countries or smaller regions within a single country, conducted with 
regular frequency (e.g.\ weekly), public health agencies hope to detect such 
outbreaks and identify the affected areas in a timely manner 
\citep{Salmon2016,Cakici2010}. One 
group of methods utilized for this purpose are \emph{scan statistics}. 
The aim of this paper is to introduce a novel scan statistic for 
space-time data, prove its effectiveness relative to alternatives, and provide a 
reference to an implementation of the proposed method.

Scan statistics are used to detect clusters in spatial and spatio-temporal 
datasets, and have been applied in fields as diverse as 
astronomy \citep{Bidin2009}, ecology \citep{Tuia2008}, and criminology 
\citep{Minamisava2009}. The primary use of scan statistics continues to be in 
public health surveillance, however, and this is the area of application we will
focus on to demonstrate the method proposed in this paper.
Scan statistics detect clusters by first \emph{scanning} over regions of 
interest, calculating \emph{statistics} for all potential clusters. The scan 
statistic is then obtained as the maximum of these statistics, and the 
corresponding \emph{most likely cluster} (MLC) thus identified. To determine 
whether such a cluster was detected as a result of an emerging outbreak or 
because of chance occurrence, formal hypothesis testing can be conducted. 
Typically, $P$-values or rejection regions are obtained through the use of Monte 
Carlo methods. 

A common assumption for many scan statistics, such as those proposed by
\citet{Kulldorff1997a}, \citet{Gangnon2001}, \citet{Tango2005}, and 
\citet{Li2011}, is that the observed number of events in a given area and time 
period follows a Poisson distribution. This distribution is a natural first 
choice for rare event data or for count data with a large or unclear upper 
bound. However, scan statistics that rely on the Poisson assumption have been 
criticized on two points in particular: their inability to capture 
overdispersion in the data \citep{Tango2011,deLima2015} and their inability to 
handle zero-inflated counts \citep{Cancado2014,deLima2015}. Additionally, a 
debate has recently flared up around the use of traditional scan statistics, 
such as those implemented in the software \textsc{SaTScan}™ \citep{satscan}, and 
the hypothesis testing procedures used when carrying out prospective analyzes
\citep{Correa2015a,Kulldorff2015,Correa2015b}. Essentially, the issue 
discussed is if and how both past and future analyzes should be accounted for 
when deciding which significance level to use for the analysis at hand. In the 
latest contribution to the debate, \citet{Tango2016} argues that 
\citet{Correa2015a} are right but on the wrong grounds. Reiterating the point 
made in \citet{Tango2011}, he asserts that prospective analyzes should not be 
conducted using conditional expected counts, i.e.\ conditional on the total 
number of observed cases. Rather, all baseline parameters should be 
estimated independently of the current study period, on past data that is 
believed to be free from disease outbreaks and other anomalies. In the 
nomenclature of \citet{Neill2006a}, \citet{Tango2016} argues that prospective 
analysis should be conducted with \emph{expectation-based} scan statistics, not 
\emph{population-based} scan statistics.

Motivated by the need for expectation-based scan statistics and the inability of
many current scan statistics to handle excess zeros, we take inspiration from
the work of \citet{Cancado2014} and propose an expectation-based prospective 
scan statistic for spatio-temporal counts governed by a zero-inflated Poisson (ZIP) 
distribution. The proposed scan statistic along with several others are made 
available in the open-source \textsc{R} package \verb|scanstatistics|, which can 
be seen as a complement to the \textsc{SaTScan}™ \citep{satscan} software and 
its \textsc{R} interface \verb|rsatscan| \citep{rsatscan}, as well as to more 
general disease surveillance packages such as \verb|surveillance| 
\citep{Salmon2016a}.

\noindent
The structure of this paper is as follows: In Section \ref{sec:methods},
the theoretical basis and assumptions of an outbreak model based on the ZIP 
distribution is described, and algorithms for parameter estimation and 
calculation of the proposed scan statistic are given. 
Section \ref{sec:sim} then describes a simulation study conducted to test the 
performance of the method relative to another scan statistic. This is
followed by an application on Campylobacter data in Section \ref{sec:application}.
Finally, conclusions and major takeaways are given in Section 
\ref{sec:conclusion}.

\section{Methods} \label{sec:methods}
Counts of zero are not rare in applications, and can come about in 
several ways. In disease surveillance, for example, zero cases may be reported 
for a given area and time period because none of the ill individuals decided to 
visit the doctor, and this chance circumstances is in some sense responsible for 
these cases remaining undetected.  We refer to such zero counts as 
\emph{sampling zeros}.
Quite a different reason to see a count of zero is that no disease was present, 
and assuming no false positives, consequently no cases were recorded. 
This latter type of observation, which cannot be anything but zero, will be 
referred to as a \emph{structural zero}. A more deceptive variant of a 
structural zero is one that occurs because of a continual failure to look for 
the disease in question.
Indeed, until healthcare practitioners become aware of a possible emerging 
outbreak, such zeros may be common because no one is on the lookout for that
particular disease. Depending on its rarity, 
cases may be misdiagnosed and thus lead to falsely reported zeros. In that case, 
it may be advisable for health authorities to treat zero counts with caution.
However they come about, the ability to account for 
structural zeros in addition to sampling zeros can enhance disease surveillance. 
In Section \ref{sec:zip}, we will first present the zero-inflated Poisson (ZIP) 
distribution, which indeed accounts for structural zeros, followed by a model 
for disease outbreaks based on the ZIP distribution in Section 
\ref{sec:outbreak-model}. Section \ref{sec:likelihood} then presents the details 
for parameter inference needed for the expectation-based ZIP scan statistic 
presented in Section \ref{sec:zipscanstat}.

\subsection{The zero-inflated Poisson distribution} \label{sec:zip}
The zero-inflated Poisson distribution is a mixture between a point mass at zero
and a Poisson distribution. If the probability of drawing a zero from the point
mass, i.e.\ a structural zero, is denoted $p$, and the
Poisson distribution is parametrized by its expected value $\mu$, a random
variable $Y$ following a ZIP distribution has probability mass function (pmf)
\citep[p.\ 193]{Johnson2005}

\begin{align} \label{eq:ZIPdist}
\P(Y = y; p, \mu) = 
\begin{cases}
p + (1 - p) e^{-\mu}, \quad y = 0 \\
(1 - p) \mu^y e^{-\mu} / y!, \quad ~~y = 1, 2, \ldots
\end{cases}
\end{align}
The random variable $Y \sim \text{ZIP}(p, \mu)$ then has expected value 
$\E[Y] = (1-p) \mu$ and variance $\V(Y) = (1-p) \mu + p (1-p) \mu^2$. The 
variance is seen to be greater than the mean, and the ZIP distribution is thus 
able to account for overdispersion caused by excess zeros, relative to the 
ordinary Poisson distribution. The next section shows how the ZIP 
distribution can be used to model outbreaks involving structural zeros.

\subsection{A Space-Time ZIP Outbreak Model} \label{sec:outbreak-model}
We consider a scenario in which disease cases 
$Y_{it} \sim \text{ZIP}(p_{it}, \mu_{it})$ are reported at regular intervals of 
time $t \in \mathbb{Z}$ from locations enumerated $i=1, \ldots, n$. 
The intervals of time may for example be hours, days or weeks, and the locations 
correspond to regions within which counts are aggregated. For example, a 
location may be a district in a country, and all individual disease cases that 
are diagnosed in that district within say a week are reported for the district as 
a whole. The observed data for the time period under surveillance consists of 
counts $\{y_{it}\}$, where $i=1,\ldots,n$ and $t = 1, \ldots, T$.
In our notation, time is counted backwards, so that $t=1$ denotes the most 
recent time interval and $T$ the interval at the start of the study period. 
Additionally, we suppose that enough data from past non-outbreak conditions 
exists for us to reliably estimate the parameters $(p_{it}, \mu_{it})$ of the 
ZIP distribution. This is a separate task from the calculation of the scan 
statistic, and may involve fitting a ZIP regression model using e.g.\ the EM 
algorithm \citep{Lambert1992}. 
The global null hypothesis is 
that no disease outbreak is occurring in any of the locations $i=1,2,\ldots,n$ 
during the study period $t=1,2,\ldots,T$, which translates to the statement that 
the counts $Y_{it}$ are independently ZIP-distributed with the 
parameters $(p_{it}, \mu_{it})$ estimated from the historical data. Formally, 
\begin{align*}
H_0 : ~ Y_{it} \sim \text{ZIP}(p_{it}, \mu_{it}) 
\quad \text{for }  i=1,\ldots,n \, \text{  and  } \, t=1,\ldots,T.
\end{align*}
If an outbreak occurs, we suppose that this happens according to a so called
\emph{hotspot} model \citep[see e.g.][]{Tango2011}, in which the Poisson 
expected value parameter $\mu_{it}$ is increased by a factor $q_W > 1$ (the 
\emph{relative risk}) for counts inside a given space-time window 
$W = (Z, D)$. Here, the \emph{zone} $Z$ is a collection of locations 
representing the spatial extent of the cluster, and $D$ similarly
designates the cluster's temporal component.

For such a window $W$, the alternative hypothesis may be stated
\begin{align*}
H_1(W) : ~ Y_{it} \sim
\begin{cases}
\text{ZIP}(p_{it}, q_W \mu_{it}), \quad (i, t) \in W \\
\text{ZIP}(p_{it}, \mu_{it}), \quad ~~~~(i, t) \not\in W .
\end{cases} 
\end{align*}
The problem is that the window $W$ is unknown: we know neither the 
duration of the outbreak nor which locations are affected. However, because we 
are only interested in detecting ongoing outbreaks, we limit the
investigation to windows $W$ with temporal components $D = (1, 2, \ldots, d)$
stretching from the present ($t=1$) to $d \leq T$ units into the past, where
$T$ is the maximum outbreak duration we are willing to consider.
Similarly, we are mainly interested in detecting 
\emph{localized} outbreaks, for which the affected locations are close to one 
another spatially. This restriction can be achieved by investigating only a 
pre-defined set of zones $\mathcal{Z}$ fulfilling such a condition. Given these 
zones, we have a collection $\mathcal{W}$ of space-time windows $W$ to test the 
above alternative hypothesis on.

One way of constructing the set of zones $\mathcal{Z}$ is 
to consider each location and its $k$ nearest neighbors ($k$-NN) as a zone, for 
$k=0,1,\ldots,K_{\max}$. Here, $K_{\max}$ may be determined by not letting any 
zone include more than half of all locations, for example. This set can be 
enriched, for example by considering connected subsets of these $k$-NN
zones \citep{Tango2005}. Other approaches are possible, both those specified 
independently of the response data, and those created by a more data-driven 
approach \citep[as in][]{Neill2012}. To determine the proximity of locations, 
distances can be calculated between them. For example, if locations correspond 
to districts, then the geographical distance between the largest cities in each
of two districts can serve as a distance measurement.

\begin{figure}[H]
  \centering
    \includegraphics[width=0.4\textwidth]{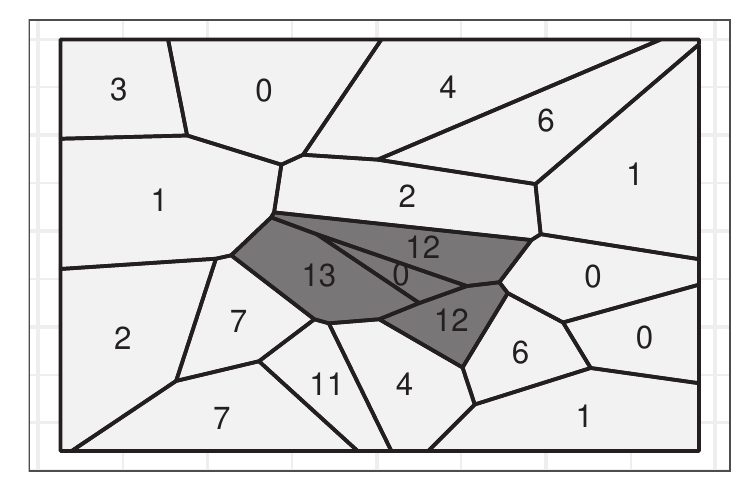} 
  \caption{Example of an outbreak zone (dark shade) containing a structural zero in the center location, at a given timepoint. The numbers in the figure indicate the observed count in a given time period.} \label{fig:middle_zero_fig}
\end{figure}

To see the use of a scan statistic for the above outbreak model, consider the
outbreak in the set of locations corresponding to the darkly shaded regions in 
Figure \ref{fig:middle_zero_fig}. 
If the expanding $k$-nearest neighbor method 
described above is used to construct the set of potential outbreak zones, no 
zone capturing the true outbreak locations could avoid capturing the middle 
location with a structural zero reported for the current time period.
Zones like this one would be penalized by scan statistics that do not account 
for structural zeros, such as those based on a simple Poisson model for counts. 
In contrast, a zero count will be a natural part of an outbreak cluster for a
scan statistic based on the ZIP outbreak model above, and therefore not 
penalized. In the next section, we construct such a scan statistic based on the 
likelihood function implied by the model.

\subsubsection{Likelihood Specification} \label{sec:likelihood}
The assumption that the parameters $(p_{it}, \mu_{it})$ are known, but in 
practice replaced by point estimates $(\hat{p}_{it}, \hat{\mu}_{it})$ based on 
past data, means that the likelihood function is constant under the null 
hypothesis. Under the 
alternative hypothesis of an outbreak in a space-time window $W$, the likelihood 
$L(q_W | \{\mu_{it}\}, \{p_{it}\})$
is a function of the relative risk $q_W$. Unlike the case for an ordinary 
Poisson distribution, such as that in \citet{Neill2005}, there is no analytical 
solution for the likelihood-maximizing value of $q_W$ for the zero-inflated 
Poisson outbreak model. A solution can instead be obtained numerically, e.g.\ by
an expectation-maximization (EM) algorithm \citep{Dempster1977}. 
A similar algorithm was given for a population-based ZIP scan statistic in 
\citet{Cancado2014}, which was formulated for a different setting and outbreak 
model; these differences will be clarified after the definition of the scan 
statistic in Section \ref{sec:zipscanstat}. To derive an 
EM algorithm for estimating $q_W$, we will first take a look at the complete
data likelihood function, i.e.\ when it is known which zeros are structural 
zeros.
To this end, let $\delta_{it}$ be an indicator variable that takes the 
value $1$ if the count at location $i$ and time $t$ is a structural zero, and 
the value $0$ if it is not. By the description of the zero-inflated Poisson 
distribution given previously, it follows that $\P(\delta_{it} = 1) = p_{it}$, 
and that the conditional pmf for a count at location $i$ 
and time $t$ is described by the probabilities
\begin{align*}
    \P(Y_{it} = y_{it} | \delta_{it} = 0) 
      &= \exp\left( -q_W \mu_{it} 
             \right) 
         (q_W \mu_{it})^{y_{it}} / y_{it}!, 
         \quad y_{it} = 0, 1, \ldots,
\end{align*}
and with $\P(Y_{it} = 0 | \delta_{it} = 1) = 1$.
This holds under the alternative hypothesis $H_1(W)$ for a given space-time 
window $W$. For locations and times outside of $W$, or under the null 
hypothesis, the corresponding probabilities are obtained by setting $q_W = 1$ 
above. Supposing that the values of the structural zero indicators are known,
we can express the complete information likelihood corresponding to the 
alternative hypothesis $H_1(W)$ as
\begin{align} \label{eq:EMlikelihood}
L(q_W | \{\mu_{it}\}, \{p_{it}\}, \{\delta_{it}\})
&\propto
\exp\left( 
  \sum_{(i,t) \in W} (1 - \delta_{it}) \left[y_{it} \log(q_W) - q_W \mu_{it}\right] 
\right).
\end{align}
In a practice, the structural zero indicators $\delta_{it}$ may not be known, 
but they can be estimated by their conditional expected values, 
provided the parameter $q_W$ is known. 
Vice versa, the analytical maximum likelihood estimate of $q_W > 1$ can easily 
be calculated if the structural zero indicators are available. This suggests 
that an iterative solution for the maximum likelihood estimate:
For each space-time window $W$, first initialize the relative risk $q_W$ at its 
value under the null hypothesis, $q_W^{(0)} = 1$. Then for $k \geq 1$, repeat 
the following steps until convergence of the (incomplete information) likelihood 
$L(q_W | \{\mu_{it}\}, \{p_{it}\})$:
\begin{itemize}
\item \textbf{Expectation step}: for all pairs $(i, t) \in W$ 
  for which $y_{it} = 0$, set
  \begin{align} \label{eq:eststep}
  \hat{\delta}_{it}^{(k)}
  =
  \frac{p_{it}}{p_{it} + (1-p_{it}) \exp(-\hat{q}_W^{(k-1)} \mu_{it})}.
  \end{align}
  For those $(i, t) \in W$ for which $y_{it} > 0$, set $\hat{\delta}_{it} = 0$ 
  at the first iteration and leave this unchanged.
\item \textbf{Maximization step}: set
  \begin{align} \label{eq:maxstep}
  \hat{q}_W^{(k)}
  =
  \max \left\{1, 
              \frac{\sum_{(i,t) \in W}   y_{it} }
                  {\sum_{(i,t) \in W} \mu_{it} (1 - \hat{\delta}_{it}^{(k)})} 
       \right\},
  \end{align}
  and increase $k$ by 1.
\end{itemize}
For a predetermined $\epsilon > 0$,
the relative convergence criterion $|L(\hat{q}_W^{(k)}) / L(\hat{q}_W^{(k-1)}) - 1| < \epsilon$
is used to decide when to terminate the iterations, at which point the optimal 
value $\hat{q}_W$ is available. The right hand side of Equation 
\eqref{eq:maxstep} shows that observations that are likely to be structural 
zeros contributes positively to the relative risk estimate by reducing the 
influence of the corresponding Poisson mean parameter $\mu_{it}$ in the 
denominator; this would not be the case in the standard Poisson outbreak model.
%

For each space-time window $W$, a log-likelihood ratio statistic is then
calculated as
\begin{align} \label{eq:loglihoodstat}
\lambda_W 
&= 
\log 
\left(
\frac
{L(\hat{q}_W | \{\hat{p}_{it}\}, \{ \hat{\mu}_{it} \})}
{L(1 | \{\hat{p}_{it}\}, \{ \hat{\mu}_{it} \})} 
\right)
=
\log 
\left(
\frac
{\prod_{i, t} \P(Y_{it} = y_{it}; \hat{p}_{it}, \hat{q}_W \hat{\mu}_{it})}
{\prod_{i, t} \P(Y_{it} = y_{it}; \hat{p}_{it}, \hat{\mu}_{it})}
\right).
\end{align}

\subsection{The Expectation-based ZIP Scan Statistic} \label{sec:zipscanstat}
Because the cardinality of $\mathcal{W}$ can be in the range of thousands or 
hundreds of thousands for typical applications, a multiple testing problem would 
arise if each hypothesis test was conducted without adjustment for the other 
tests. On the other hand, a Bonferroni-type correction for the significance 
level would mean that only very rare events would be detected \citep{Abdi2007}. 
This may preclude small- to medium-sized outbreaks from detection. The solution 
suggested by \citet{Kulldorff1995} is to focus on the cluster $W$ which gives 
the largest value of the statistic $\lambda_W$.
The proposed scan statistic and corresponding most likely cluster (MLC) are thus
given by
\begin{align*}
  \lambda^* = \max_{W \in \mathcal{W}} \lambda_W \quad \text{and} \quad
  W^*       = \argmax_{W \in \mathcal{W}} \lambda_W.
\end{align*}
\noindent
Because the distribution of the scan statistic is unavailable in closed form, 
hypothesis testing cannot be carried out using
plug-in formulas. An often used alternative is to calculate a $P$-value by Monte 
Carlo simulation, as suggested by \citet{Kulldorff1995}. This procedure can be 
summarized as follows.
\begin{enumerate}[noitemsep]
\item Determine the scan statistic $\lambda^{*}_{\text{obs}}$ on the 
      observed data, along with the corresponding most likely cluster $W^*$.
\item Use the estimated baseline parameters 
      $(\hat{p}_{it}, \hat{\mu}_{it})$ to 
      simulate $R$ replicate data sets for all locations $i$ and times $t$ in 
      the study period.
\item Calculate a replicate scan statistic $\lambda^*_j$ on each simulated data 
      set $j=1,\ldots,R$.
\item Reject the null hypothesis of no outbreak if the Monte Carlo $P$-value,
      given by 
      \begin{align} \label{eq:Pvalue}
        P = \left(1 + \sum_{j=1}^R \IX\{ \lambda^*_j > \lambda^*_{\text{obs}} \} \right) / (1 + R)
      \end{align}
      is smaller than the chosen significance level $\alpha$.
      Here, $\IX$ is the indicator function.
\end{enumerate}
Secondary clusters $W$ with potential outbreaks can be found by looking at the
other top order statistics of $\{\lambda_W\}_{W \in \mathcal{W}}$ and comparing
their values to the simulated scan statistics. Such tests are conservative, 
because, e.g., the third largest observed value will be compared to the (first) 
largest simulated value of $\lambda_W$ \citep{Kulldorff1997a}. Typically, 
$R = 999$ replications are made, which means 
that the computational cost is increased almost a thousandfold over calculating 
the statistic on just the observed data. As proposed by \citet{Abrams2006}, this 
cost can be reduced significantly by simulating only a smaller number of scan 
statistics and fitting a Gumbel distribution to these replicates. A $P$-value 
can then be calculated as the tail probability of the observed scan statistic 
for the fitted distribution. Another possibility is to use \emph{empirical $P$-values},
obtained by applying Equation \ref{eq:Pvalue} to previously calculated 
values of the scan statistic, rather than simulated values. This approach may
be more reliable in practice, because hypothesis testing using Monte Carlo 
$P$-values have been shown to be miscalibrated for observational data, typically 
requiring a much lower significance level than the nominal $\alpha$ to achieve 
the target level of false positives \citep{Neill2009a}. 

The scan statistic proposed above (call it EB-ZIP) differs from that given by 
\citet{Cancado2014} (PB-ZIP) in three major ways:
\begin{enumerate}[noitemsep]
  \item Analyses conducted with the PB-ZIP statistic are conditional on the 
        total observed count: to conduct hypothesis testing by Monte Carlo 
        replication, new counts are simulated conditional on that their total
        equals the total seen in the original data. As shown by 
        \citet{Tango2011}, this conditioning can in a worst-case scenario ensure 
        that no outbreak is detected, and may in general have low power to 
        detect outbreaks that affect a large proportion of the area under 
        surveillance \citep{Neill2009a} . On the other hand, the proposed EB-ZIP 
        statistic does not condition on the observed total count when conducting
        hypothesis testing.
  \item The PB-ZIP statistic uses the populations at risk for each area directly 
        in the calculation of the scan statistic, while the relative risk and 
        zero probability parameters are equal for all locations inside or 
        outside a given cluster in the alternative hypothesis (and the same for
        all locations under the null hypothesis). The populations at risk may 
        not always be available or even applicable, and can be harder to adjust
        for e.g.\ seasonal or day-of-week effects compared to the expected 
        values $\mu_{it}$ \citep{Neill2006a}. Indeed, with a sufficient amount
        of historical data, the parameters $(p_{it}, \mu_{it})$ can easily be
        estimated in a regression-type framework.      
  \item The PB-ZIP statistic was introduced for a purely spatial outbreak 
        detection scenario, which may be adequate for retrospective analyses. 
        The method presented in this paper is designed for space-time data, 
        suitable for prospective surveillance situations in which an outbreak 
        may have started to emerge only during the few most recent time periods.
\end{enumerate}
\noindent In brief, using the nomenclature introducted by \citet{Neill2006a}, the 
scan statistic presented in this paper is \emph{expectation-based}, rather than
\emph{population-based}.
In the next 
section, we conduct a simulation study to investigate the detection abilities of 
the proposed scan statistic in a prospective space-time setting.

\section{Simulation Study} \label{sec:sim}
A comprehensive simulation study was performed to test the detection timeliness
and spatial accuracy of the proposed scan statistic (called EB-ZIP below). 
A smaller subset is illustrated below, but the conclusions put forth are drawn
from the full set of simulations. 
For comparison, the expectation-based Poisson 
\citep[EB-POI;][]{Neill2005} and the 
population-based Poisson \citep[PB-POI;][]{Kulldorff2001} were also calculated. 
Following an initial period of 9 weeks with ZIP-distributed data simulated from 
non-outbreak conditions at $n=100$ randomly placed locations, outbreaks with
different relative risks were simulated for 11 consecutive weeks at specified 
subsets of the locations. 
Thus the temporal window of each scan statistic covered an increasing amount of
outbreak data for each new week scanned. The maximum 
outbreak duration considered was $T = 10$ weeks, which is the same maximum 
duration used in Section \ref{sec:application}.
An outbreak was defined as detected in the first week for which the calculated 
$P$-value fell below a given significance level $\alpha$. A range of different
values of $\alpha$ were tried, in order to evaluate the sensitivity of the 
detection timeliness, spatial accuracy and false positive rate with respect to 
the significance level.
The zones used in the analysis were constructed by the 
nearest-neighbor method described in Section \ref{sec:outbreak-model}, and 
contained at most 25 locations each. Multiple outbreak \emph{scenarios} were 
considered, defined by all possible combinations of the following parameters:
\begin{itemize}[noitemsep,topsep=0pt]
  \item The baseline ZIP count parameter: $\mu_{it} = \mu \in \{1, 5, 10\}$.
  \item The baseline structural zero probability: $p_{it} = p \in \{0.01, 0.05, 0.15, 0.25, 0.5\}$.
  \item The outbreak relative risk: $q_W \in \{1, 1.1, 1.25, 1.5, 2\}$.
  \item The spatial extent of the outbreak: either 5 or 20 locations.
\end{itemize}
The value $q=1$ corresponds to a non-outbreak and serves as a check of the 
false positive rate of each method.
For each outbreak scenario, 1000 outbreaks were simulated and scan statistics
calculated on each. Additionally, 999 non-outbreaks were 
simulated for each set of baseline parameters $(p, \mu)$, to allow for 
$P$-value calculations and hypothesis testing. Because the difference between 
Monte Carlo and Monte Carlo + Gumbel $P$-values was deemed to be sufficiently 
small, only the latter type of $P$-value was used in the analysis.

To evaluate the spatial accuracy of the ZIP scan statistic, we measured the 
spatial precision $S_{\text{P}}$ and the spatial recall 
$S_{\text{R}}$ as defined by \citet[p.\ 356]{Neill2012}. Denoting by $Z^*$ the 
spatial component (\emph{zone}) of the detected space-time cluster, and 
$Z^{\text{true}}$ the true outbreak zone, these measures are given by:
\begin{align*}
S_{\text{P}} = \frac{|Z^* \cap Z^{\text{true}}|}
                     {|Z^*|}, \quad
S_{\text{R}} = \frac{|Z^* \cap Z^{\text{true}}|}
                     {|Z^{\text{true}}|}, 
\end{align*}
where, for example, $|Z^*|$ denotes the number of locations contained in the 
zone $Z^*$.
The spatial precision informs us about the relevance of the detected cluster,
in terms of what fraction of the detected locations truly are affected by an 
outbreak. The recall, on the other hand, tells us what fraction of the locations
truly affected by an outbreak that we captured---how well does the method avoid
false negatives? There is a trade-off between precision and recall, because 
adding more locations to the reported cluster will eventually improve the 
recall, but penalize the precision. Conversely, removing locations from the 
reported cluster will eventually improve precision at the expense of recall, 
given that at least one true outbreak location is captured. As a summary of both 
precision and recall, the harmonic mean 
$F = (S_{\text{P}}^{-1} + S_{\text{P}}^{-1})^{-1}$ of the two measures can be used
\citep{Neill2009b}. 
We computed these measures for each simulated outbreak, noting in 
particular their values at the first day for which the $P$-value fell below the 
significance level $\alpha$.

\subsection{Simulation Results} \label{sec:sim-results}
\begin{figure}[H]
  \centering
  \includegraphics{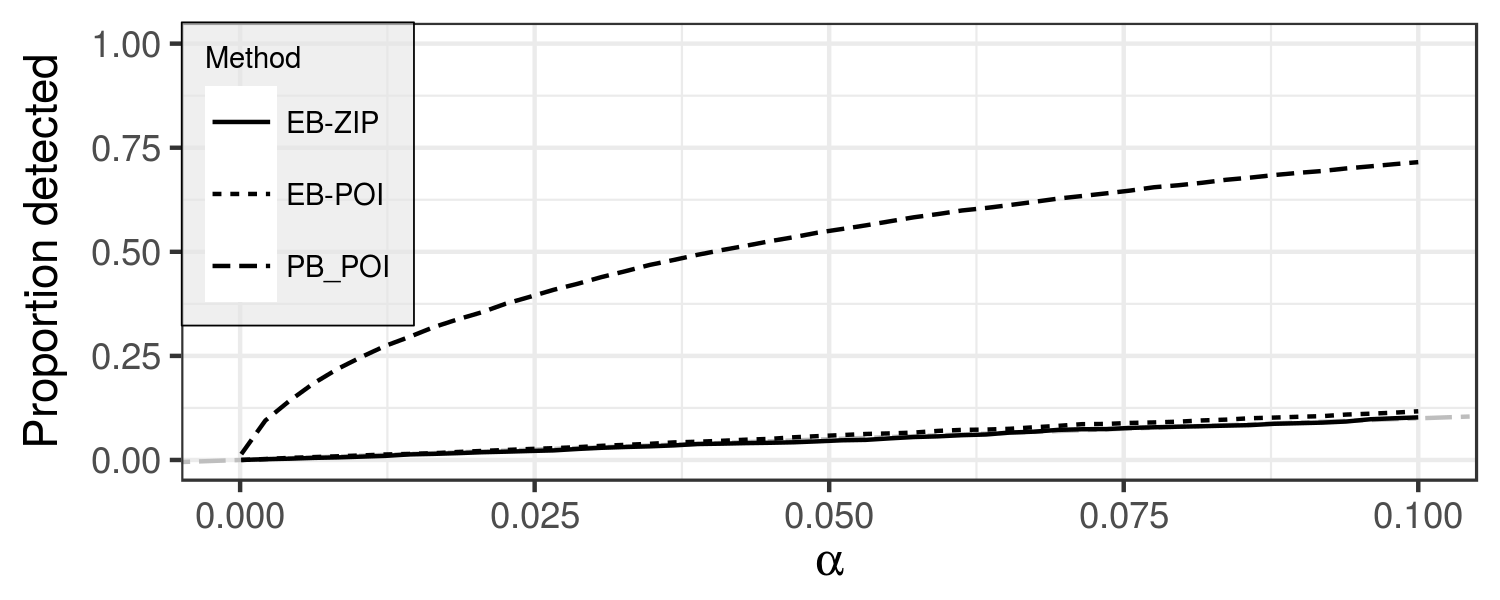} 
  \captionsetup{format=plain}
  \caption{Proportion of outbreaks detected versus significance level for non-outbreak data. The grey dot-dashed line marks the relationship $y=x$. The figure corresponds to the scenario with parameters $p=0.15$ and $\mu=5$.} \label{fig:nonobsim}
\end{figure}

Before measuring the timeliness of a detection method, it is informative to 
first investigate its false positive rate. A method that signals a detection 
regardless of whether an outbreak is actually occuring will have great 
timeliness but is utterly useless in practice.
We simulate 1000 non-outbreaks ($q_W = 1$) for each combination of baseline 
parameters $(p, \mu)$, and compute the proportion of false positives 
($P < \alpha$) for a range of $\alpha$ values.
In Figure \ref{fig:nonobsim} we show the proportion of outbreaks detected by 
each scan statistic, with the significance level $\alpha$ ranging between 0.0001 
and 0.1. The plot corresponds to the scenario with parameters $\mu = 5$ and 
$p=0.15$. 
%

\noindent
At a given significance level $\alpha$, a detection method should on expectation 
preferably not give a higher proportion of false alarms than $\alpha$, which is 
the proportion expected if the assumptions on which the method is based are true. 
Indeed, this is the case for the proposed EB-ZIP method, and Figure
\ref{fig:nonobsim} shows that the line for EB-ZIP (solid black) follows the line 
$y=x$ (dot-dashed grey) very well. Also clear from Figure \ref{fig:nonobsim} is 
that the PB-POI method signals far more outbreaks than is expected, while the 
EB-POI method only signals slightly more outbreaks than EB-ZIP. For other
parameter settings, not shown, it can be concluded that the proportion of false 
alarms of PB-POI increases rapidly with the structural zero probability $p$ and 
decreases moderately with $\mu$, although it is seemingly always higher than the 
nominal $\alpha$. For the EB-POI method, the false alarm ratio is mostly lower
than $\alpha$, but the opposite is true for high values of $p$.

\begin{figure}[H]
    \centering
    \begin{subfigure}[t]{0.5\textwidth}
        \centering
        \includegraphics[]{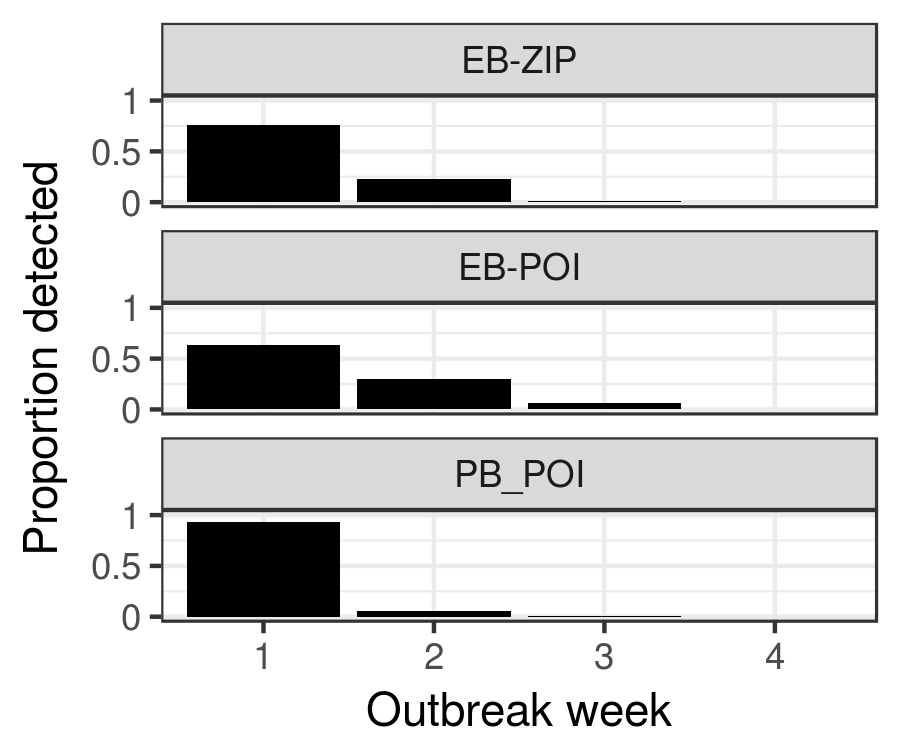}
        \caption{} \label{fig:weeklyprop}
    \end{subfigure}%
    %
    \begin{subfigure}[t]{0.5\textwidth}
        \centering
        \includegraphics[]{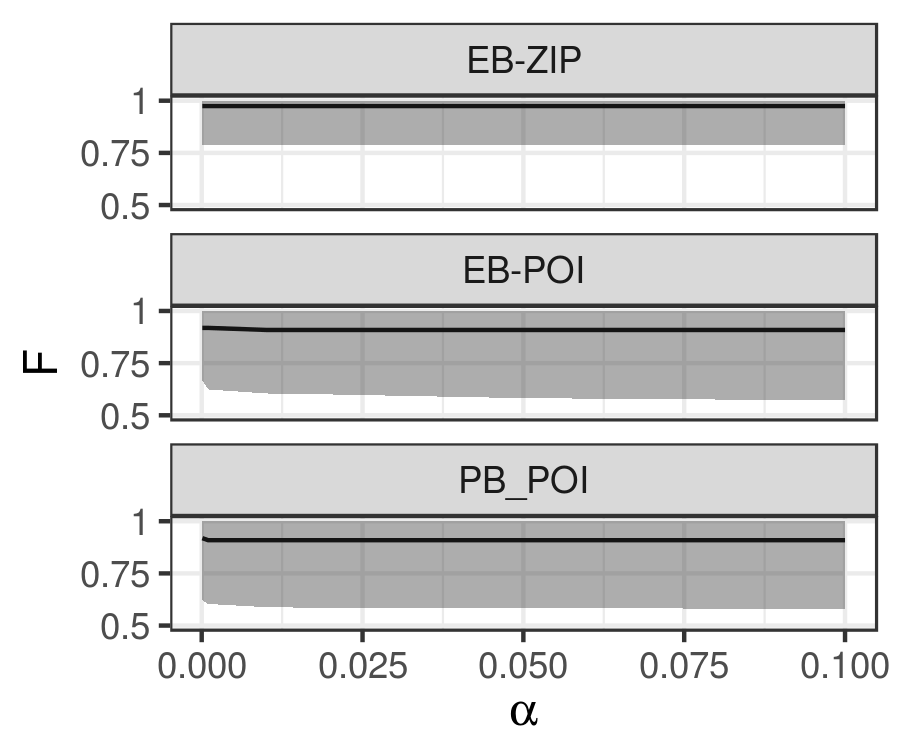}
        \caption{} \label{fig:F1alpha}
    \end{subfigure}
    \caption{In (a), the proportion of outbreaks detected per week, for a fixed
             significance level $\alpha = 1/50$. In (b), the solid line marks
             the median harmonic mean $F$ in outbreak week 3 for detected outbreaks 
             ($P < \alpha$) for different significance levels $\alpha$. The
             shaded regions are 90\% pointwise confidence intervals for $F$.
             Parameter settings were $p=0.15$, $\mu=5$, $q_W = 1.5$, and 
             the outbreak affected 20 out of 100 locations.} \label{fig:detectime}
\end{figure}

The detection timeliness of a method that does not give too many false alarms 
may still be irrelevant if the clusters detected do not overlap with the actual
locations affected by a disease outbreak. We therefore investigate the spatial
accuracy of detected outbreaks at different significance levels $\alpha$.
In Figure \ref{fig:detectime}\subref{fig:weeklyprop}, we show the proportion of 
outbreaks detected per week of the outbreak for each of the three scan 
statistics, for a fixed significance level $\alpha = 0.05$. 
\noindent
In this instance, 
most outbreaks have been detected by week 3 for each scan statistic.
In Figure \ref{fig:detectime}\subref{fig:F1alpha} the harmonic mean $F$ 
defined previously is shown for varying levels of $\alpha$ for outbreaks
detected in week 3.
Clearly, the PB-POI scan statistic detects outbreaks early---even when there is
not outbreak, as in Figure \ref{fig:nonobsim}---but the 90\% confidence interval
for $F$ stretches just below 0.6 to 1 for most significance levels,
meaning that its spatial accuracy is somewhat erratic. The EB-ZIP scan statistic also
detects outbreaks early---most of them in week 1 for these parameter settings---but
matches this timeliness with a high and steady spatial accuracy. 

\begin{figure}[H]
  \centering
  \includegraphics{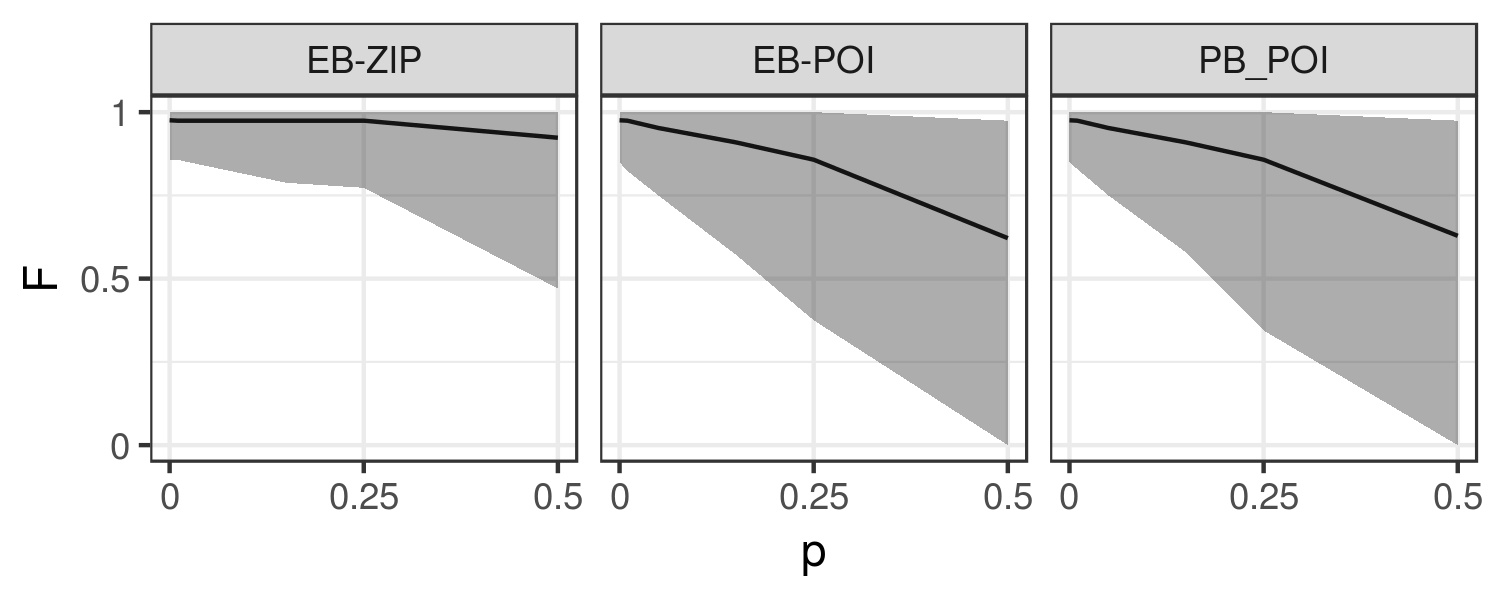} 
  \captionsetup{format=plain}
  \caption{Median (solid line) and 90\% pointwise confidence interval (shaded region) of the harmonic mean $F$ in week 3 of the outbreak, for different values of the structural zero probability $p$. Parameter settings were $\mu=5$, $q_W = 1.5$, and the outbreak affected 20 out of 100 locations.} \label{fig:limp}
\end{figure}

Lastly, we examine the performance of the ZIP scan statistic as the structural
zero probability approaches zero. At this limit, the ZIP distribution becomes 
the Poisson distribution and the performance of the EB-ZIP and PB-ZIP methods 
should converge. Figure \ref{fig:limp} shows that the 
harmonic mean $F$ of all three scan statistics converge to nearly 1 when $p \to 0$;
the solid line indicates the median of $F$ over simulations, and the shaded region
stretches from the 5th to the 95th percentile of $F$.
Looking at this from the other direction, it is clear that the relative 
performance of the ZIP scan statistic, in terms of spatial accuracy, becomes 
better the larger the structural zero probability $p$ is. The EB-POI and PB-POI
methods also show more variation in $F$ as $p$ increases, relative to EB-ZIP.
\vspace{-10pt}
\noindent
The overall results of the simulation study indicate that the expectation-based ZIP
scan statistic is able to accurately detect outbreaks in a timely manner,
particularly when non-zero counts are large on average, or when structural zeros 
are abundant. These qualities are improved when the spatial size of the outbreak 
is large or when the magnitude of the outbreak, in terms of the relative risk 
$q_W$, is significant. To demonstrate what is required to apply the proposed 
scan statistic on real disease data, and how it may differ in the clusters
reported compared to the two other scan statistics considered in this section,
we next apply the ZIP scan statistic to the weekly cases of Campylobacteriosis 
in the districts of Germany.

\section{Application: Campylobacteriosis in Germany} \label{sec:application}
\begin{figure}[H]
    \centering
    \begin{subfigure}[t]{0.4\textwidth}
        \centering
        \includegraphics[scale=0.6]{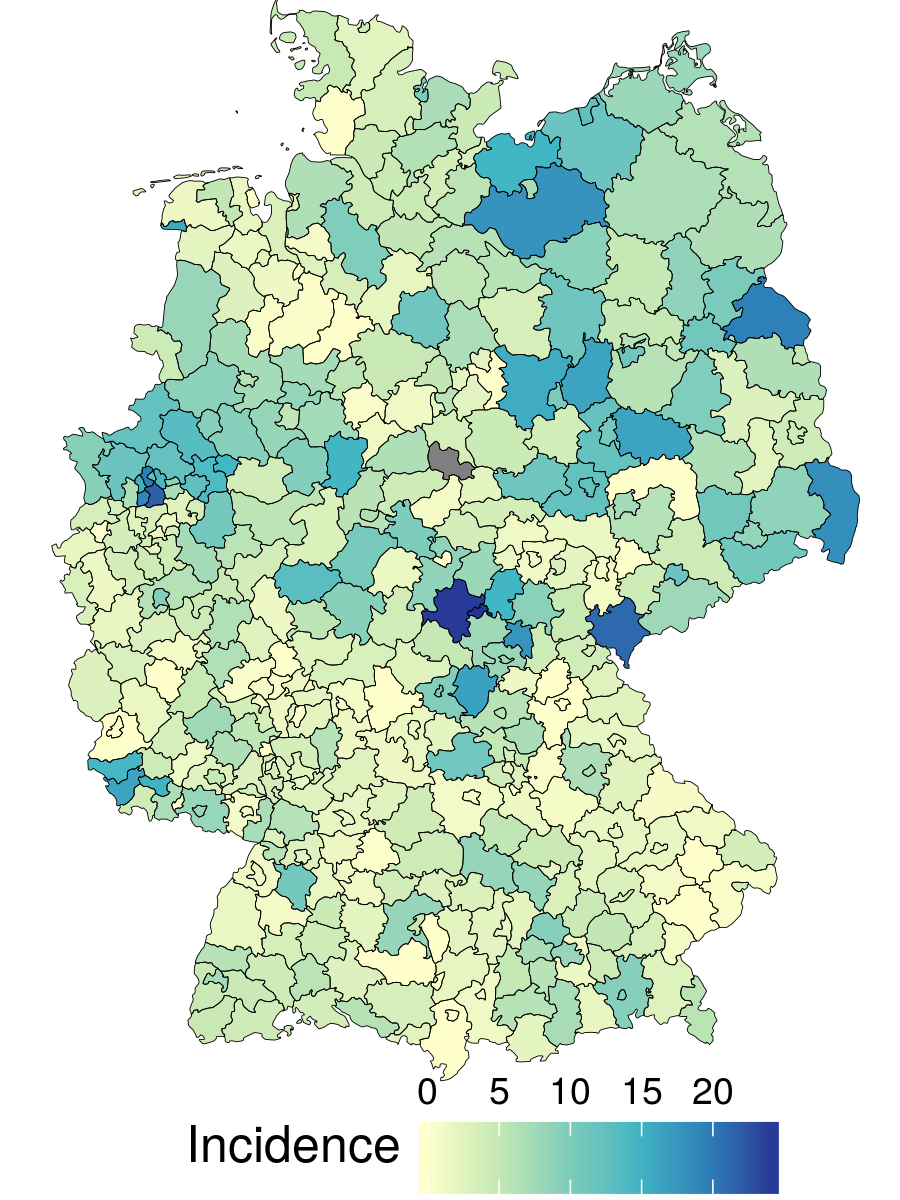}
        \caption{} \label{fig:kreis}
    \end{subfigure}%
    \hspace*{-0.9em}
    \begin{subfigure}[t]{0.6\textwidth}
        \centering
        \includegraphics[scale=0.9]{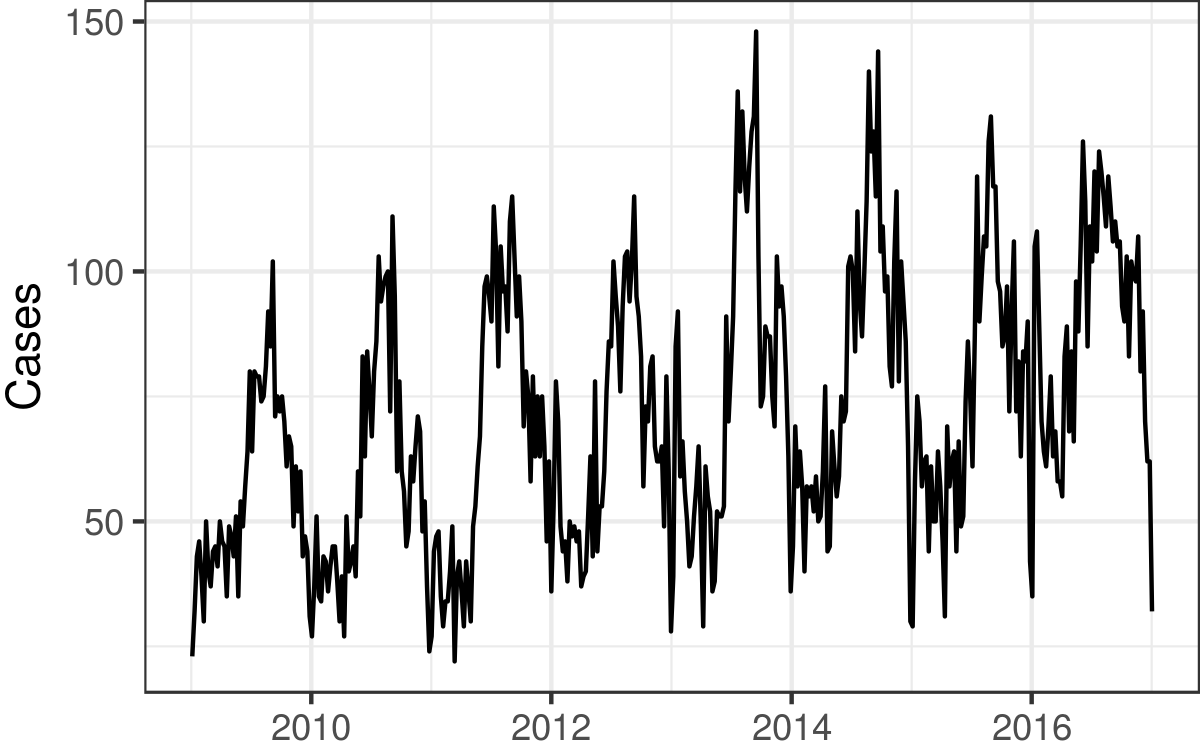}
        \caption{} \label{fig:germany_sums}
    \end{subfigure}
    \caption{In \emph{(a)}, the incidence (new cases per 100,000 people) of Campylobacteriosis (\emph{Campylobacter coli}) in the districts of Germany in 2016. In \emph{(b)}, the total number of cases per week in Germany is shown for the period 2009--2016.} \label{fig:map_counts}
\end{figure}

\defcitealias{survstat}{RKI, 2017}
\emph{Campylobacteriosis} is a diarrhoeal disease caused by the bacteria
\emph{Campylobacter}, with approximately 80--90\% of human cases in 
industrialized countries attributed to the species \emph{Campylobacter jejuni} 
and 5--10\% attributed to \emph{Campylobacter coli} \citep{CamBookChap6,CamBookChap12}.
The main route of transmission to humans is via poorly
cooked meat and other meat products---particularly poultry---but also via raw
or contaminated milk, and water or ice.
Campylobacteriosis has one of the highest incidences among the notifiable 
diseases monitored by the Robert Koch Institute (RKI) in Germany. 
From the public database \emph{SurvStat@RKI 2.0} \citepalias{survstat}, we collected 
the weekly number of cases of Campylobacteriosis due to \emph{Campylobacter coli}
in the years 2011--2016, for each of the 402 districts (\emph{kreise}) of 
Germany, subdividing the counts for Berlin by the city's 12 boroughs. A map of 
the 402 with districts shaded by incidence of Campylobacteriosis in 2016 is 
shown in Figure \ref{fig:kreis}. In Figure \ref{fig:germany_sums}, the weekly counts 
of Campylobacteriosis aggregated to the whole of Germany are shown for the period
2009--2016. A clear seasonal pattern with peaks in the summer months is visible
here, which is not the case for the corresponding district-level plot (not shown).
%
%
%
\vspace{-10pt}
\defcitealias{SuWeNa2013}{Südwestfalen-Nachrichten, 2013}       
\noindent
Noticeable outliers in the years 2011--2015 were matched against news articles 
and epidemiological records from approximately the same times and places. One
such outlier was found to correspond to an outbreak in the district Märkischer 
Kreis in
January of 2013 \citepalias{SuWeNa2013}; it was replaced by the median of counts in 
the surrounding weeks from all the baseline years. Data from the period 
2011--2015 were then used as a baseline on which a zero-inflated regression model 
with was fitted by maximum likelihood. The parameters of the model were 
specified as
\begin{align*}
  \log(\mu_{it}) &= \log(\text{pop}_{it}) + \alpha_{\mu} 
                  + \gamma_{\mu} \delta_i^{\text{urban}} 
                  + \theta_{\mu}^{\text{state}(i)}
                  + \beta_0 \tilde{t} + f_{\mu}(\text{east}_i, \text{north}_i) \\
                 &+ \beta_1 \sin \left( \frac{2 \pi \text{ISOweek}_t}{52.18} + \phi_1 \right)
                  + \beta_2 \sin \left( \frac{4 \pi \text{ISOweek}_t}{52.18} + \phi_2 \right) \\
  \text{logit}(p_{it})
    &= \log(\text{pop}_{it}) + \alpha_{p} 
     + \gamma_{p} \delta_i^{\text{urban}} 
     + \theta_{p}^{\text{state}(i)}
     + f_{p}(\text{east}_i, \text{north}_i),
\end{align*}
where $\tilde{t}$ is a standardized measure of time;
$\delta_i^{\text{urban}}$ is an indicator variable taking the value 1 if
district $i$ is classified as an urban district (\emph{stadtkreis}) as opposed
to a rural district; 
$\theta_{\boldsymbol{\cdot}}^{\text{state}(i)}$ is a coefficient shared by all districts
in the same federal state as district $i$;
$\text{pop}_{it}$ is the estimated 
population of district $i$ at time $t$ (used as an offset); the two sine terms
representing seasonal fluctuations over the course of a year, using a period
of 52.18 weeks instead of 52 to account for leap weeks; and $f_{\mu}$ and
$f_{p}$ are smooth functions of the coordinates of each location, represented
by thin plate regression spline bases \citep{Wood2003} with a basis dimension
of 150 in each case. The form of the model and the basis dimensions were arrived
at by minimization of AIC over a finite set of models using functions from the 
\textsc{R} packages \verb|pscl| \citep{pscl} and \verb|mgcv| \citep{Wood2003}. 
A Poisson GLM was also fit with the same form of $\mu_{it}$ as above, to use for 
the expectation-based Poisson scan statistic.
For our analysis, we considered outbreaks with a maximum duration of 10 weeks,
which was the maximum reporting delay found to be relevant by 
\citet{Salmon2015a}, and the investigated spatial zones had the flexible shape 
proposed by \citet{Tango2005} with at most 10 districts in each zone. 
This yielded a total of $53,\!830$ spatial zones and thus ten 
times as many potential space-time clusters. For hypothesis testing, we calculated 
scan statistics for each week of the baseline data and used these values in 
Equation \eqref{eq:Pvalue} to obtain empirical $P$-values, for reasons discussed in 
Section \ref{sec:zipscanstat}. These were calculated using a 10 week rolling 
window and the same spatial zones used for the test period.
Unfortunately, because the population-based Poisson scan statistic
conducts its analysis based on the value of the total observed count, there are
too few values of the statistic in the baseline period to match those in the 
study period. Therefore, Monte Carlo $P$-values had to be used for this method.

\begin{figure}[H]
  \centering
    \includegraphics[trim=0cm 0.5cm 0cm 0.5cm, scale=0.6]{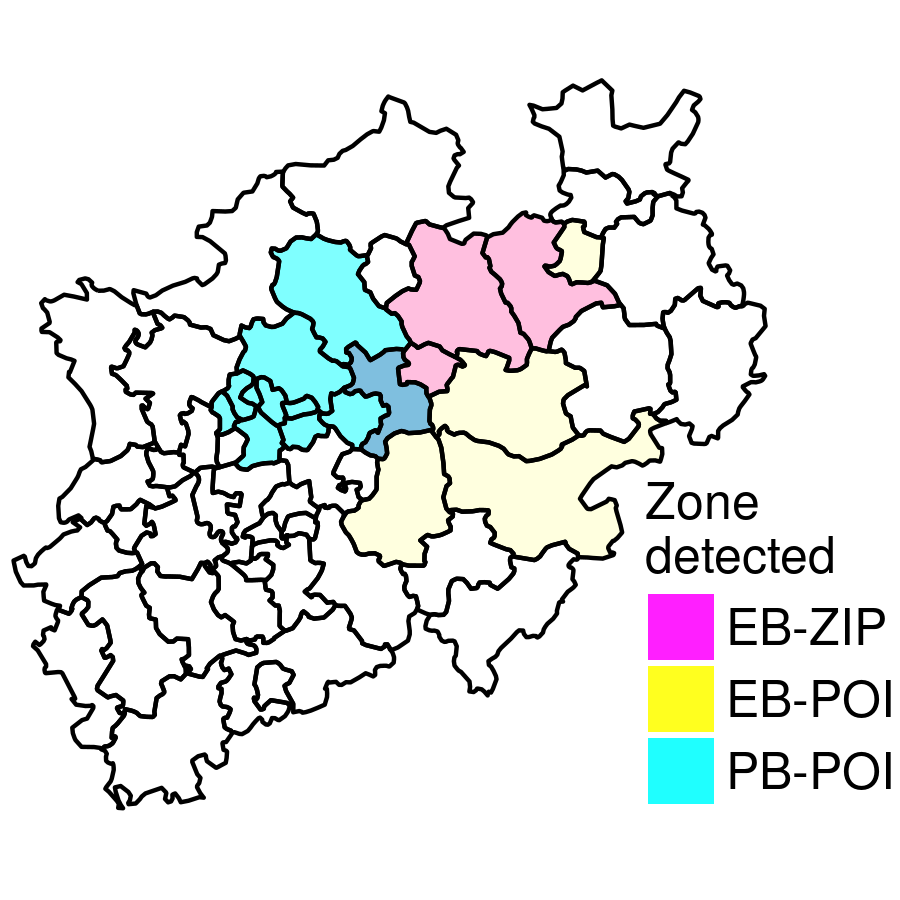} 
  \caption{Map of the districts of North Rhine-Westphalia, in which the zones
           reported by each scan statistic in week 12 of 2016 are colored.} \label{fig:cluster_map}
\end{figure}

Data from 2016 were analyzed using the proposed expectation-based
ZIP statistic (EB-ZIP), the expectation-based Poisson statistic (EB-POI) and the
population-based Poisson statistic (PB-POI). This yielded scan statistics 
calculated in weeks 9--51 of that year. The PB-POI scan statistic 
consistently reported very low $P$-values ($<10^{-4}$) throughout the period; 
this tendency to signal a detection (for common nominal significance levels) was 
also seen in the simulation study. The two expectation-based scan statistics, on 
the other hand, generally report low $P$-values ($<0.05$) in weeks 9--31, and 
mostly higher values in the following period. These scan statistics also tend
to find clusters that are smaller in the spatial dimension (median of 5 
locations for EB-ZIP; 6 for EB-POI) than PB-POI (median 10 locations, which is
the maximum). In Figure \ref{fig:cluster_map}, we show the spatial component
(zones) of the most likely cluster reported by each scan statistic in week 12
of 2016. All these zones lie in the state of North Rhine-Westphalia, although 
only one district, Unna, is reported by all three methods. The remainder of the
10 districts reported by the PB-ZIP method lie to the west of Unna, while those
of the EB-ZIP method lie more to the north-east and are fully engulfed by the
8 districs reported by the EB-POI method. The $P$-values reported by the
EB-ZIP, EB-POI and PB-POI methods are approximately 0.07, 0.1 and $2 \cdot 10^{-5}$,
respectively. Although the number of cases in week 12 were fairly low overall,
multiple districts captured in the cluster of each statistic had shown counts
of 3 and above in the previous few weeks, something not seen to the same extent
in the same period in earlier years.
\noindent
Reported outbreak durations are more similar however, with a median reported duration of
10 weeks for EB-POI and PB-POI, and 9 weeks for EB-ZIP.
Perhaps of more interest is the degree of overlap between the spatial components 
of the reported most likely clusters of each method. Defining the spatial 
overlap as the fraction of locations in common out of the total reported for a 
pair of the methods, the average overlap between the EB-ZIP and EB-POI methods 
was 61\%, while only 26\% between EB-ZIP and PB-POI. Between EB-POI
and PB-POI, it was 34\%. Upon closer inspection, it seems that all clusters
reported by the PB-POI method lies in the state of North Rhine-Westphalia,
which is the most populous state of Germany. 
The EB-ZIP method, on the other hand, reports clusters across 7 states, many in
the west, south-west and south of Germany, and the EB-POI method reports clusters across
5 states with similar geographical characteristics. In terms of outbreaks, this
paints a different picture than the incidence rates shown in Figure 
\ref{fig:kreis}, which are generally high in the north-east and low in the south
and south-west. Despite the low $P$-values reported by the three scan statistics 
in many of the weeks of the study period, no `significant' result could be matched to
events described in news articles and epidemiological records for the same time
period. As a check, we also ran the three scan statistics on the single-district outbreak 
in January of 2013 mentioned earlier. This outbreak was found in the MLCs reported
by the EB-ZIP and EB-POI methods, but not that of the PB-POI method.
The lack of data on confirmed clusters is a situation frequently met in evaluating 
new detection methods. For example, in disease surveillance a more systematic investigation 
of cases using decision support methods is often needed, but this does not 
always result in data directly amenable for cluster analysis.
Nevertheless, the proposed method and accompanying R package could be integrated
into such a decision support system to help guide the search efforts and ensure that 
no important outbreaks are missed.

\section{Conclusions} \label{sec:conclusion}
An excess of zero counts in public health data may cause scan statistics based
on the Poisson distribution to perform suboptimally. In addition, many 
traditional scan statistics estimate baseline parameters from the very data that
they attempt to detect outbreaks in. Critics argue that these parameters ought
instead to be estimated on historical data that does not overlap with the data 
under investigation, and that analyses made should not be conditional on the
observed total count \citep{Tango2011,Tango2016}.
In this paper, we have proposed an expectation-based scan statistic for
zero-inflated Poisson data that addresses these issues. By way of simulation,
we have shown that the method performs well under a range of conditions.
In particular, the proposed ZIP scan statistic performs well relative to two
scan statistics based on a simple Poisson model, under conditions in which 
non-zero counts are distant from zero, or in which structural zeros are abundant. 
The detection power, both in terms of spatial accuracy and timeliness of 
detection, improves with the size and length of the outbreak, as well as its 
effect on raising counts above their baseline. To demonstrate the scan statistic,
we applied it to weekly cases of Campylobacteriosis in the districts of Germany. 
This analysis served to illustrate some of the qualitative differences
of the results reported by the proposed method and two alternative scan statistics.

Apart from an excess of zeros, real-world data is often characterized by a
greater variation in the non-zero counts than would be expected under the
assumption of Poisson- or ZIP-distributed counts. Recent work by \citet{deLima2015}
considers population-based scan statistics for the overdispersed Poisson and
double Poisson distributions. A possible extension of the work in this paper
may thus be to derive the corresponding expectation-based scan statistics for
these two distributions. The methods used in this paper are provided in
the free and open-source \textsc{R} package \verb|scanstatistics|, available
from CRAN \citep{scanstatistics}.

\section*{Acknowledgements} 
The authors thank Martin Sköld for constructive feedback during the writing
of this article.
The authors also acknowledge financial support from the Swedish Research Council 
(Grant 2013:05204, Grant 2015-05182\_VR).

\bibliographystyle{apalike}
\bibliography{ms}
\end{document}